# Giant spin Hall effect in half-Heusler alloy topological semimetal YPtBi grown at low temperature

Takanori Shirokura[1] and Pham Nam Hai[1,2]*

*[1] Department of Electrical and Electronic Engineering, Tokyo Institute of Technology,*

*Meguro, Tokyo 152-8550, Japan*

*[2] Center for Spintronics Research Network (CSRN), The University of Tokyo,*

*Bunkyo, Tokyo 113-8656, Japan*

*Corresponding author: pham.n.ab@m.titech.ac.jp

**Abstract**

Half-Heusler alloy topological semimetal YPtBi is a promising candidate for an efficient spin source material having both large spin Hall angle $\theta_{SH}$ and high thermal stability. However, high-quality YPtBi thin films with low bulk carrier density are usually grown at 600°C, which exceeds the limitation of 400°C for back end of line (BEOL) process. Here, we investigate the crystallinity and spin Hall effect of YPtBi thin films grown at lower growth temperature down to 300°C. Although $\theta_{SH}$ degraded with lowering the growth temperature to 300°C due to degradation of the crystallinity, it was recovered by reducing the sputtering Ar gas pressure. We achieved a giant $\theta_{SH}$ up to 8.2 and demonstrated efficient spin-orbit torque magnetization switching by ultralow current density of ~$10^5$ A/cm$^2$ in YPtBi grown at 300°C with the Ar gas pressure of 1 Pa. Our results provide the recipe to achieve giant $\theta_{SH}$ in YPtBi grown at lower growth temperature suitable for BEOL process.

**Introduction**

Spin orbit torque (SOT) generated by the spin Hall effect in a non-magnetic layer can be used for manipulating the magnetization of an adjacent magnetic layer,[1] which is promising for spintronic devices such as magnetoresistive random access memory,[1,2] spin torque oscillator,[3,4] and racetrack memory.[5] Numerous studies on efficient spin source materials have been conducted with focus on improving the charge-to-spin conversion efficiency, or spin Hall angle $\theta_{SH}$.[6,7,8,9,10,11,12] Early works show that heavy metals (HMs) with strong spin orbit interaction have moderately large $\theta_{SH}$ of 0.08-0.3 and good compatibility with back end of line (BEOL) processes,[6,7,8] and thus, the first-generation SOT devices have been developed with HMs.[13,14] Recently, chalcogenide-based topological insulators (TIs) have been shown to have very large $\theta_{SH}$ (>1) thanks to their large Berry phase originated from the Dirac-like dispersion of their topological surface states (TSSs).[9,10,11,12] However, utilizing the chalcogenide based TIs in realistic SOT devices is still challenging due to their pour compatibility with BEOL processes such as low thermal stability and high toxicity, because they consist of mostly V group (Bi,Sb) and VI group (Se,Te) elements.[15,16] Therefore, the second-generation SOT devices based on chalcogenide-based TIs have not yet been adopted by the industry despite their attracting feature of large $\theta_{SH}$.[17]

Recently, we have demonstrated large $\theta_{SH}$ in YPtBi,[18] which is a half-Heulser alloy topological semimetals with no toxic element.[19,20,21] We have shown that YPtBi has not only large effective spin Hall angle $\theta_{SH}^{eff}$ up to 4.1 but also high thermal stability of 600°C, which is favorable for BEOL process. In YPtBi, suppression of bulk carrier or bulk conductivity is a key to achieve large $\theta_{\text{SH}}^{\text{eff}}$.[18,22] Since YPtBi is basically a zero-gap topological insulator, the bulk carriers are generated by crystal defects. Therefore, previous reports adopted high growth temperature of 600°C during sputtering to improve the crystal quality.[18,22] However, such a high growth temperature exceeds the limitation of 400°C of BEOL process. For fabrication of realistic SOT devices on top of silicon electronics during BEOL process, it is essential to reduce the growth temperature to below 400°C.

In this work, we investigate the crystallinity and spin Hall effect of YPtBi grown at lower temperature down to 300°C. We found that $\theta_{\text{SH}}^{\text{eff}}$ and the effective spin Hall conductivity $\sigma_{\text{SH}}^{\text{eff}}$ of YPtBi decrease with lowering the growth temperature from $\theta_{\text{SH}}^{\text{eff}} = 2.8$ and $\sigma_{\text{SH}}^{\text{eff}} = 0.89 \times 10^5$ $(\hbar/2e)\Omega^{-1}\text{m}^{-1}$ at 600°C to $\theta_{\text{SH}}^{\text{eff}} = 1.2$ and $\sigma_{\text{SH}}^{\text{eff}} = 0.33 \times 10^5$ $(\hbar/2e)\Omega^{-1}\text{m}^{-1}$ at 300°C due to degradation of the crystal quality. On the other hand, we found that by decreasing the sputtering Ar gas pressure, $\sigma_{\text{SH}}^{\text{eff}}$ can be recovered by improving the crystallinity thanks to higher migration energy of sputtered atoms. We realized a giant $\theta_{\text{SH}}^{\text{eff}}$ up to 8.2, and then performed SOT magnetization switching by ultralow current density in the YPtBi layer grown at 300°C with the Ar gas pressure of 1 Pa. Our results open up a way to achieve giant $\theta_{\text{SH}}^{\text{eff}}$ in YPtBi grown at low growth temperature suitable for BEOL process.

**Effect of growth temperature on the crystallinity and spin Hall effect of YPtBi**

First, we investigated how the growth temperature ($T_{\text{G}}$) affects the crystallinity and spin Hall effect of YPtBi. We prepared 30 nm-thick YPtBi films grown at $T_{\text{G}} = 300, 400, 500$ and 600°C by co-sputtering Y, Pt, and Bi targets. The Ar gas pressure during YPtBi sputtering was fixed to 2.0 Pa, and the atomic Y/Pt composition ratio was fixed to the exact stoichiometry of 1.0 to minimize the bulk carriers.[22] The sample structure was Ta (1.0) / MgAl$_2$O$_4$ (5.0) / YPtBi (30) / c-Sapphire (the number are the layer thicknesses in nm). Here, we assumed that the 1 nm-thick Ta layer on MgAl$_2$O$_4$ was fully oxidized after exposure to the air. Figure 1(a) shows the X-ray diffraction (XRD) spectra for the YPtBi samples grown at various temperature. Peaks from YPtBi(111) orientations were clearly observed in all samples. Peaks from residual Bi (filled triangles) were also observed at $T_G$ below 400°C, when the Bi de-absorption rate from the YPtBi films is lower. Figure 1(b) shows the $T_G$ dependence of the integrated peak intensity of YPtBi(111) with the powder ring correction.[23] Although the integrated peak intensity is nearly unchanged at $T_G \geq$ 400°C, it dropped at $T_G$ = 300°C indicating degradation of the crystallinity. Figure 1(c) shows the $T_G$ dependence of the hole carrier density ($p$) and mobility ($\mu$) measured by using the Van der Pauw method at room temperature. $p$ and $\mu$ are nearly unchanged at $T_G \geq$ 400°C and comparable to those reported in another sputtered Bi-based half Heusler alloy.[24] However, $p$ increased to $10^{21}$ cm$^{-3}$ while $\mu$ dropped to 7 cm$^2$/Vs at $T_G$ = 300°C, reflecting degradation of the crystallinity. Therefore, weaker spin Hall effect is expected for the sample grown at 300°C due to its poor crystallinity.

We then investigate the relationship between the growth temperature and the spin Hall effect. For this purpose, we fabricated heterostructure samples consisting of Ta (1.0) / MgAl$_2$O$_4$ (1.0) / Pt (0.8) / Co (0.8) / Pt (0.8) / YPtBi (10) / c-Sapphire, where the YPtBi layer was grown at 300, 400, 500 and 600°C, and the Pt/Co/Pt layers were grown at room temperature. Below, we will refer Pt (0.8) / Co (0.8) / Pt (0.8) as CoPt (2.4) for short. Here, the parasitic spin Hall effect from the two Pt layers is negligible because the thickness of the Pt layers is few times thinner than the spin diffusion length of Pt and the layer structure of CoPt is symmetric.[25,26,27,28] After deposition, these samples were patterned into 10×60 μm² Hall bar devices with Pt / Ta electrodes for transport measurements. Figure 2(a) – (d) show the anomalous Hall resistance $R_{\text{AHE}}$ of the Hall bar devices measured with an external magnetic field $H_{\text{ext}}$ applied along the z-direction (film normal). Although all samples show perpendicular magnetic anisotropy (PMA), the squareness ratio of the hysteresis of $R_{\text{AHE}}$ becomes less than 1 at 300°C. These results suggest that the quality of CoPt on YPtBi also degrades with decreasing $T_G$. Next, we used the high-field second harmonic Hall effect measurement technique with an alternating current at 259.68 Hz for evaluation of $\theta_{\text{SH}}^{\text{eff}}$. In the high-field second harmonic Hall effect method, the second harmonic Hall resistance $R_{xy}^{2\omega}$ measured with $H_{\text{ext}}$ applied along the x-direction (parallel to the current direction) and higher than the effective perpendicular magnetic anisotropy field $H_k^{\text{eff}}$ is given by,[29,30]

$$R_{xy}^{2\omega} = \frac{R_{\text{AHE}}}{2} \frac{H_{\text{DL}}}{|H_{\text{ext}}| - H_k^{\text{eff}}} + \alpha_{\text{ONE}} |H_{\text{ext}}| + R_{\text{ANE+SSE}}, \quad (1)$$

where $H_{\text{DL}}$ is the antidamping-like field, $\alpha_{\text{ONE}}$ is a coefficient reflecting contribution from the ordinary Nernst effect, and $R_{\text{ANE+SSE}}$ is a constant reflecting contribution from the anomalous Nernst effect and the spin Seebeck effect. Figure 3(a) shows the representative high-field second harmonic Hall resistance data for the sample grown at 600°C and the corresponding fitting using Eq. (1) at bias currents of 1.8 to 3.8 mA (0.4 mA step), where the dots and solid curves are the experimental data and fitting curves, respectively. Figure 3(b) shows the relationship between the extracted values of $H_{DL}$ and the current density in the YPtBi layer ($J_{YPtBi}$) for this sample. Then, $\theta_{SH}^{eff}$ was calculated from the slope of $H_{DL}/J_{YPtBi}$ by,

$$\theta_{SH}^{eff} = \frac{2eM_S t_{CoPt}}{\hbar}\frac{H_{DL}}{J_{YPtBi}}, \quad (2)$$

where $e$ is the electron charge, $\hbar$ is the Dirac constant, $M_S$ = 431 emu/cc is the saturation magnetization of the CoPt layer measured by a superconducting quantum interference device (SQUID), and $t_{CoPt}$ = 2.4 nm is the thickness of the CoPt layer. We obtained large $\theta_{SH}^{eff}$ of 2.8 in the sample grown at 600°C thanks to the large Berry phase of TSS and its high crystallinity. By this way, we estimated $\theta_{SH}^{eff}$ and $\sigma_{SH}^{eff} = \frac{\hbar}{2e}\theta_{SH}^{eff} * \sigma_{YPtBi}$ for all samples. Here, $\sigma_{YPtBi}$ is the conductivity of the 10 nm-thick YPtBi layer, estimated from the total resistance of the heterostructures and that of a reference Ta (1.0) / MgAl$_2$O$_4$ (1.0) / Pt (0.8) / Co (0.8) / Pt (0.8) sample without the 10 nm-thick YPtBi layer. Figure 3(c) and (d) show the $T_G$ dependence of $\theta_{SH}^{eff}$ and $\sigma_{SH}^{eff}$, respectively. Both $\theta_{SH}^{eff}$ and $\sigma_{SH}^{eff}$ monotonically decreases with lowering $T_G$, even though the crystal quality is maintained for $T_G \geq 400$°C. Degradation of $\theta_{SH}^{eff}$ and $\sigma_{SH}^{eff}$ at $T_G \geq 400$°C is seemingly caused by the poor spin transparency due to degraded interface quality between CoPt and YPtBi, while further degradation of $\theta_{SH}^{eff}$ and $\sigma_{SH}^{eff}$ at $T_G$ = 300°C is caused by the poor crystallinity of YPtBi.

**Effect of sputtering Ar gas pressure on the crystallinity and spin Hall effect of YPtBi**

Next, we investigated the effect of sputtering Ar pressure ($P_{Ar}$) on the crystallinity and spin Hall effect of the sample grown at 300°C. Improvement of the bulk crystallinity of YPtBi is expected for lowering $P_{Ar}$ during YPtBi deposition, because lower $P_{Ar}$ provides higher kinetic energy of sputtered Y/Pt/Bi atoms and recoil Ar atoms, leading to higher migration energy for the atoms that arrive on the film surface.[31] Figure 4(a) shows the $P_{Ar}$ dependence of the XRD spectra for 30 nm-thick stand-alone YPtBi samples grown at 300°C. We found that YPtBi(111) orientation is maintained at lower $P_{Ar}$. Figure 4(b) shows the $P_{Ar}$ dependence of the integrated peak intensity of YPtBi(111). The intensity was improved by reducing $P_{Ar}$, indicating improvement of the crystallinity. Figure 4(c) shows the $P_{Ar}$ dependence of $p$ and $\mu$ in 30 nm-thick stand-alone YPtBi thin films measured by using the Van der Pauw method at room temperature. $p$ monotonically decreases while $\mu$ monotonically increases with lowering $P_{Ar}$, indicating improvement of their crystallinity, which is consistent with the XRD results.

Next, we investigated the spin Hall effect of YPtBi grown at 300°C with various $P_{Ar}$ in heterostructures of Ta (1.0) / MgAl$_2$O$_4$ (1.0) / Pt (0.8) / Co (0.8) / Pt (0.8) / YPtBi (10) / c-Sapphire. Figure 5(a) – (d) show $R_{AHE}$ of 10×60 μm$^2$ Hall bar devices of those samples measured with $H_{ext}$ applied along the z-direction. We found that the squareness ratio of the $R_{AHE}$ hysteresis was recovered by reducing $P_{Ar}$. Figure 6(a) – (d) show the surface morphology for these samples measured by atomic force microscopy. We observed that the grain size becomes larger as expected from the higher migration energy of atoms arrived on the surface of YPtBi at low $P_{Ar}$, consistent of the improved bulk crystal quality shown in Fig. 4 and the enhanced hysteresis of $R_{AHE}$ in Fig. 5. Figure 7(a) shows the representative high-field second harmonic Hall resistance data for the sample grown at $P_{Ar} = 1.0$ Pa with low conductivity $\sigma_{YPtBi}$ of $0.14 \times 10^5$ $\Omega^{-1}$m$^{-1}$, and the corresponding fitting using Eq. (1) at bias currents of 1.8 to 3.4 mA (0.4 mA step), where the dots and solid curves are the experimental data and fitting curves, respectively. Figure 7(b) shows the relationship between the extracted values of $H_{DL}$ and $J_{YPtBi}$ for this sample. From this slope, we obtained a giant $\theta_{SH}^{eff}$ of 8.2 thanks to suppression of bulk shunting current in this sample. This value is larger than $\theta_{SH}^{eff}$ of many conventional chalcogenide-based TIs,[9,10,11] and comparable to that of sputtered BiSb topological insulator.[32] Figure 7(c) shows the $P_{Ar}$ dependence of $\theta_{SH}^{eff}$. Although $\theta_{SH}^{eff}$ does not increase monotonously with lowering $P_{Ar}$, we achieved large $\theta_{SH}^{eff}$ of 1.5 at $P_{Ar} = 0.7$ Pa and 4.2 at $P_{Ar} = 0.5$ Pa. Figure 7(d) shows the $P_{Ar}$ dependence of $\sigma_{SH}^{eff}$. $\sigma_{SH}^{eff}$ is drastically improved at $P_{Ar} \leq 1$ Pa, and exceeds that of the sample grown at 600°C with $P_{Ar} = 2.0$ Pa (denoted by the dashed line).

Finally, we demonstrated SOT magnetization switching by ultralow current density in the sample with giant $\theta_{SH}^{eff}$ of 8.2. Figure 8(a) shows direct current (DC)-induced SOT magnetization switching with $H_{ext}$ (-0.55 ~ 0.55 kOe) applied along the $x$-direction. The switching polarity was inverted when the direction of $H_{ext}$ was reversed, consistent with the feature of SOT switching. Here, the switching polarity was same as Pt.[18,33] Figure 8(b) shows the threshold current density $J_{th}^{YPtBi}$ as a function of $H_{ext}$. Low $J_{th}^{YPtBi}$ on the order of $10^5$ Acm$^{-2}$ was achieved for entire $H_{ext}$. Figure 8(c) shows the pulse current-induced SOT switching with $H_{ext}$ of 0.18 kOe applied along the $x$-direction, where the pulse width $\tau$ was set to 50 μs ~ 10 ms. Figure 8(d) shows $J_{th}^{YPtBi}$ as a function of $\tau$. The switching current density in YPtBi is two orders of magnitude smaller than that in Pt with CoPt ferromagnetic layer at 100 ms pulse current.[34] We then fit $J_{th}^{YPtBi}$ by the thermal activation model,[35,36]

$$J_{th}^{YPtBi} = J_{th0}^{YPtBi} \left[ 1 - \frac{1}{\Delta} \ln \left( \frac{\tau}{\tau_0} \right) \right], \quad (3)$$

where, $J_{th0}^{YPtBi}$ is the threshold current density for the YPtBi layer at 0 K, $\Delta$ is the thermal stability factor, and $1/\tau_0$ (10 GHz) is the attempt frequency associated with the precession frequency of a magnetization. From the fitting, we obtained $\Delta$ = 39 and $J_{th0}^{YPtBi}$ = 7.3×10$^5$ A/cm$^2$. Finally, we demonstrate repeating SOT switching by pulse currents of $J_{YPtBi}$ = 6.8×10$^5$ A/cm$^2$ and $\tau$ = 50 μs. Figure 9(a) shows the sequence of total 105 pulse currents. Figure 9(b) shows the Hall resistance data measured with $H_{ext}$ = 0.18 kOe applied along the x-direction. We observed very robust switching by YPtBi. These results indicate that the YPtBi film grown at 300°C and low $P_{Ar}$ has SOT performance not inferior to that of the sample grown at 600°C.

**Conclusion**

We have investigated the crystallinity and spin Hall effect of YPtBi films grown at lower temperature. When we reduced the growth temperature from 600°C to 300°C, both $\theta_{SH}^{eff}$ and $\sigma_{SH}^{eff}$ decrease due to degradation of crystallinity of YPtBi and spin transparency at the interface between CoPt and YPtBi. To improve the crystallinity and spin Hall effect for samples grown at 300°C, we reduced the Ar gas pressure to increase the migration energy of atoms arrived on the surface of YPtBi.

We obtained better crystallinity of YPtBi and achieved giant $\theta_{\text{SH}}^{\text{eff}}$ up to 8.2, which is larger than those of most TIs. We demonstrated current-induced SOT magnetization switching with ultralow current density of $10^5$ A/cm$^2$ in the YPtBi layer grown at 300°C by both DC and pulse currents. These results indicate the potential of YPtBi as an efficient spin source and open up a way to achieve a giant $\theta_{\text{SH}}^{\text{eff}}$ in YPtBi grown at low growth temperature suitable for BEOL process.

**Data availability**

The data that supports the findings of this study are available from the authors upon reasonable request.

**Acknowledgements**

This work was supported by Kioxia corporation. The authors thank Tsuyoshi Kondo at Corporate R&D center, Kioxia corporation, for fruitful discussion, and S. Nakagawa Laboratory and Open Facility Center, Materials Analysis Division at Tokyo Institute of Technology, for helps on SQUID and XRD measurements. T.S. acknowledges a JSPS Fellowship (No. 21J10066).

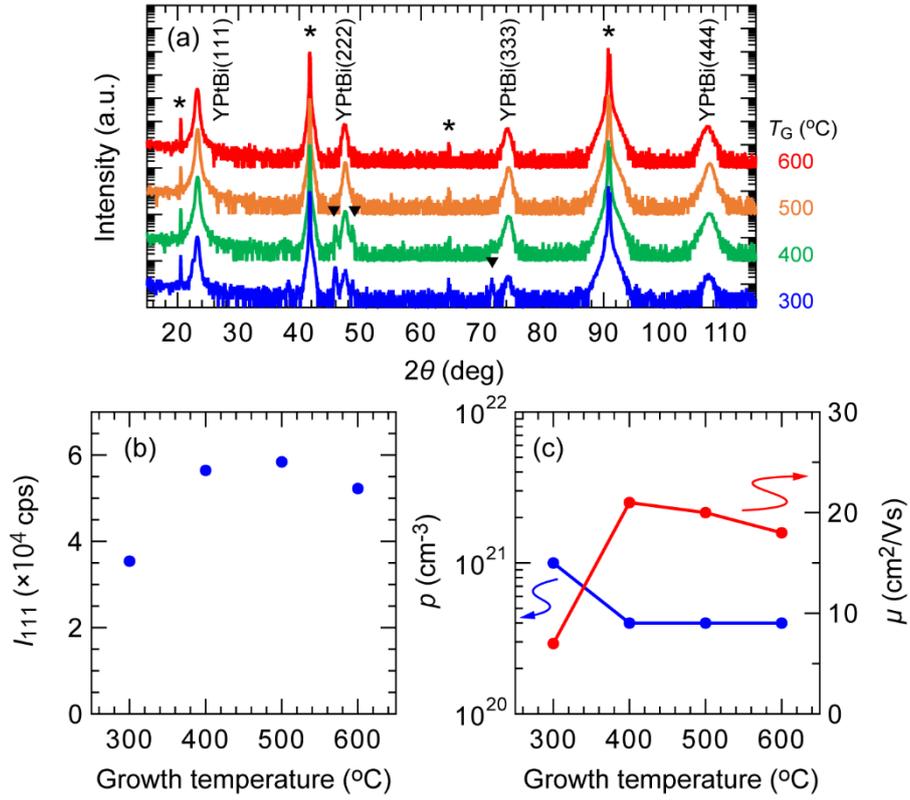

**Figure 1.** (a) XRD spectra of 30 nm-thick stand-alone YPtBi films deposited at various growth temperature ($T_G$). Asterisks denote, from left to right, the $Al_2O_3$(0003), (0006), (0009), and (00012) peaks of c-Sapphire substrate. Triangles denote, from left to right, the peaks of Bi(006), (202), and (300). (b) Growth temperature dependence of the integrated peak intensity of YPtBi(111) with the powder ring correction. (c) Growth temperature dependence of the hole carrier density $p$ and mobility $\mu$.

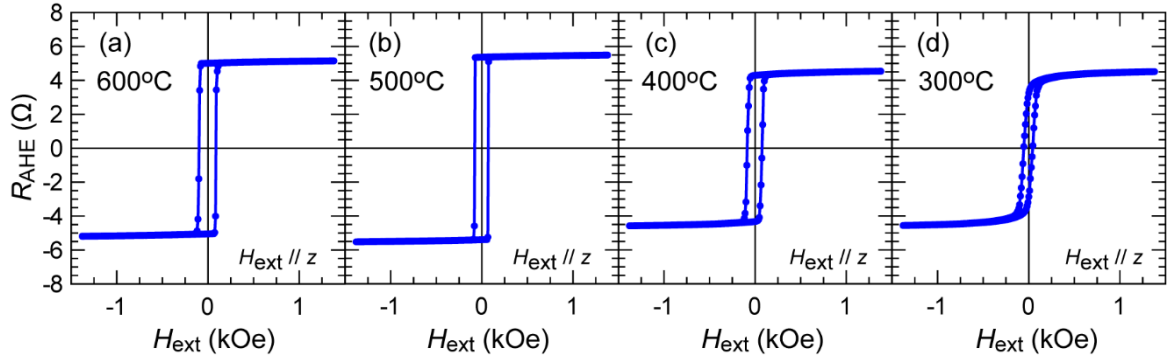

**Figure 2.** (a) – (d) Anomalous Hall resistance of 10×60 μm² Hall bar devices of Ta (1.0) / MgAl$_2$O$_4$ (1.0) / Pt (0.8) / Co (0.8) / Pt (0.8) / YPtBi (10) heterostructures, whose YPtBi layer was grown at 300, 400, 500 and 600°C, respectively.

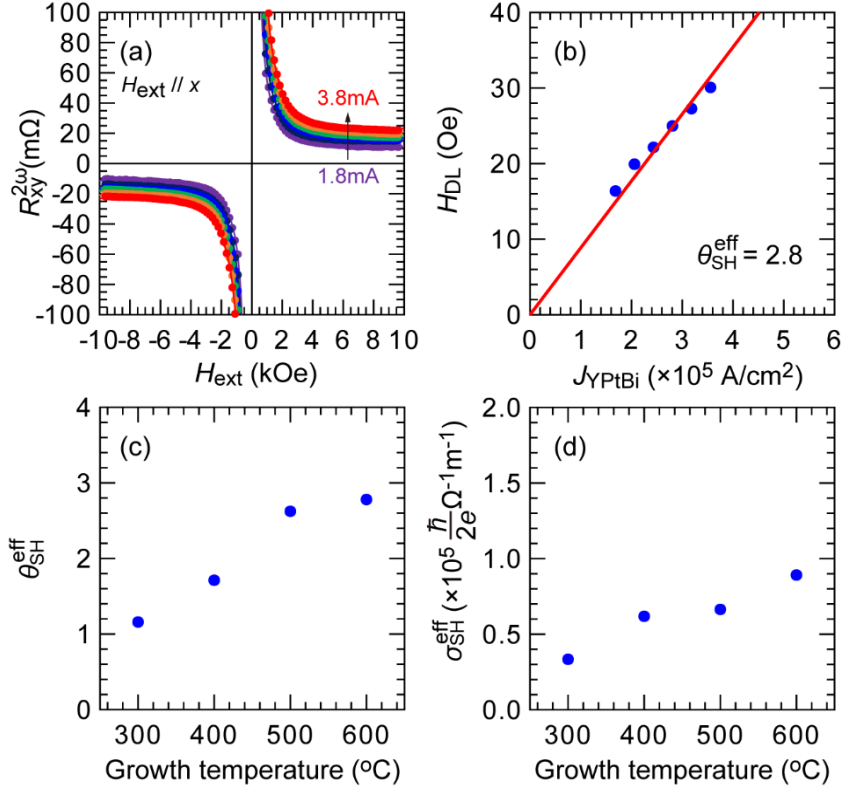

**Figure 3.** (a) Representative high-field second harmonic Hall resistance data for the heterostructure sample grown at 600°C, measured with alternating bias currents of 1.8 to 3.8 mA (0.4 mA step). Dots and solid curves are the experimental data and fitting curves using equation (1), respectively. (b) Relationship between the extracted values of the antidumping-like field ($H_{DL}$) and the current density in YPtBi ($J_{YPtBi}$) for this sample. (c), (d) Growth temperature dependence of the effective spin Hall angle $\theta_{SH}^{eff}$ and the spin Hall conductivity $\sigma_{SH}^{eff}$, respectively.

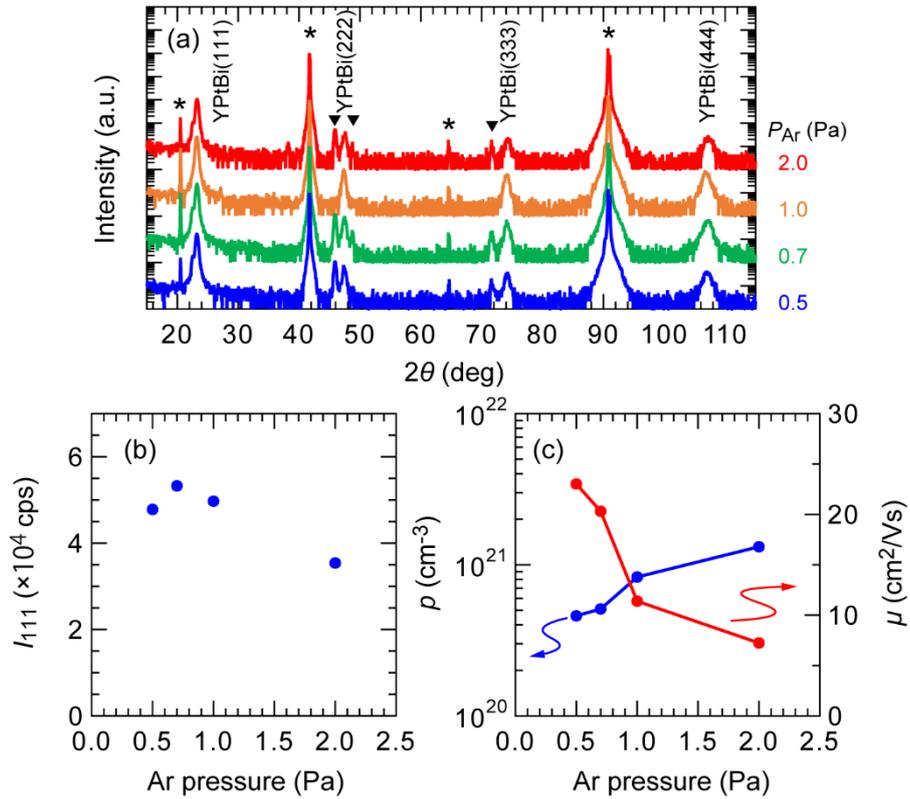

**Figure 4.** (a) XRD spectra of 30 nm-thick stand-alone YPtBi films deposited at 300°C with various Ar gas pressure ($P_{Ar}$). Asterisks denote, from left to right, the $Al_2O_3$(0003), (0006), (0009), and (00012) peaks of c-Sapphire substrate. Triangles denote, from left to right, the peaks of Bi(006), (202), and (300). (b) Ar pressure dependence of the integrated peak intensity of YPtBi(111) with the powder ring correction. (c) Ar pressure dependence of $p$ and $\mu$.

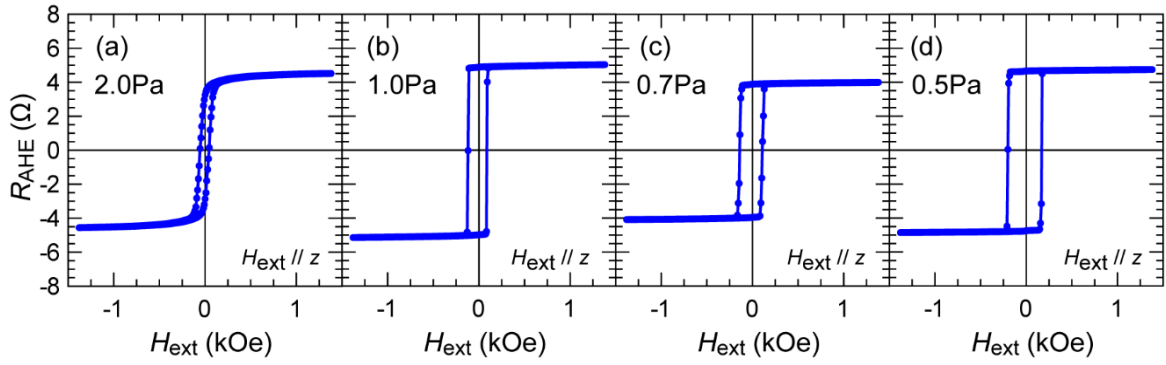

**Figure 5.** (a) – (d) Anomalous Hall resistance of 10×60 μm² Hall bar devices of Ta (1.0) / MgAl$_2$O$_4$ (1.0) / Pt (0.8) / Co (0.8) / Pt (0.8) / YPtBi (10) heterostructure samples, whose YPtBi layer was grown at 300°C with $P_{Ar}$ = 2.0, 1.0, 0.7 and 0.5 Pa, respectively.

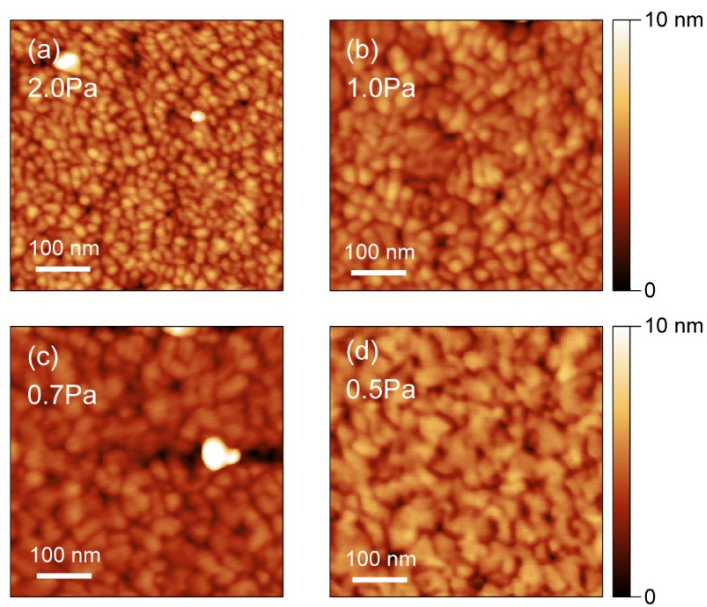

**Figure 6.** (a) – (d) Surface morphology measured by atomic force microscopy for the heterostructure samples, whose YPtBi layer was grown at 300°C with $P_{Ar}$ = 2.0, 1.0, 0.7 and 0.5 Pa, respectively.

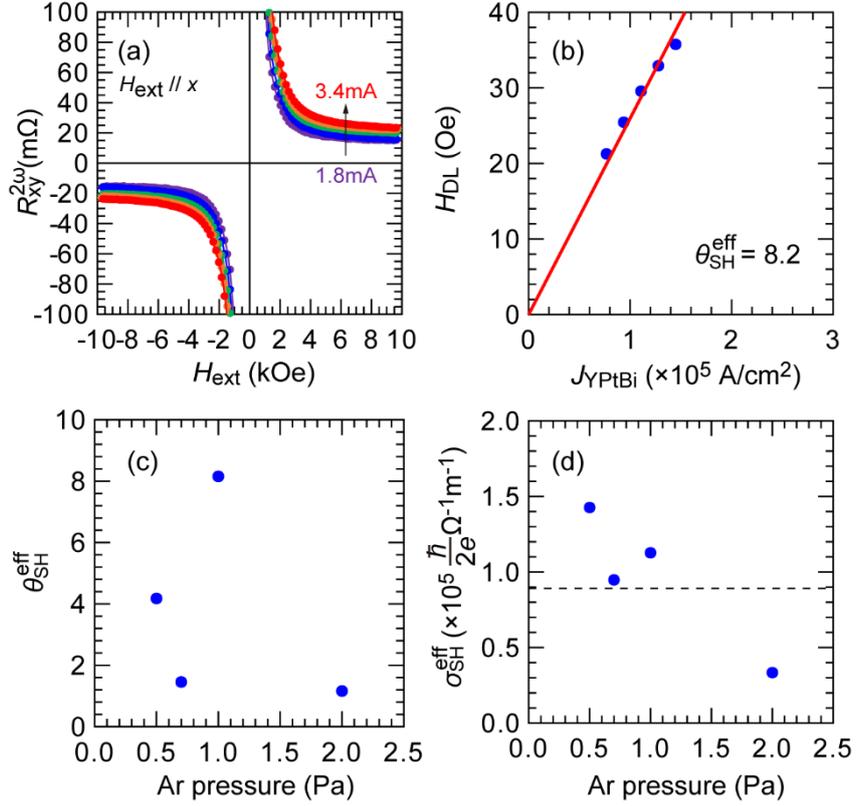

**Figure 7. (a) Representative high-field second harmonic Hall resistance data for the heterostructure sample whose YPtBi layer was grown at 300°C with $P_{Ar}$ = 1.0 Pa, measured by alternating bias currents of 1.8 to 3.4 mA (0.4 mA step). Dots and solid curves are the experimental data and fitting using equation (1), respectively. (b) Relationship between the extracted values of $H_{DL}$ and $J_{YPtBi}$ for this sample. (e) Ar pressure dependence of $\theta_{SH}^{eff}$ of heterostructure samples with YPtBi grown at 300°C. (d) Ar pressure dependence of $\sigma_{SH}^{eff}$, where the black dashed line indicates the $\sigma_{SH}^{eff}$ value of the sample grown at 600ºC with $P_{Ar}$ = 2.0 Pa.**

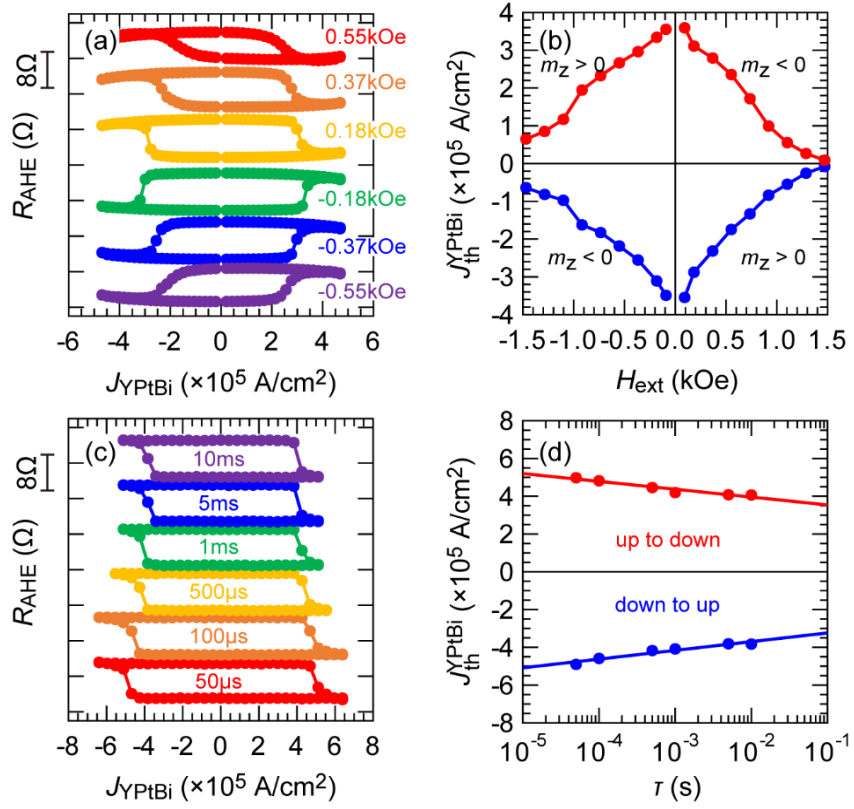

**Figure 8. Current-induced SOT magnetization switching in the heterostructure sample whose YPtBi was grown at 300°C with $P_{Ar}$ = 1.0 Pa. (a)** SOT magnetization switching by direct currents with various in-plane $H_{ext}$ (-0.55 to 0.55 kOe) applied along the *x*-direction. **(b)** Threshold switching current density in YPtBi as a function of $H_{ext}$. **(c)** SOT magnetization switching by pulse currents with various pulse width ranging from 50 μs to 10 ms and $H_{ext}$ of 0.18 kOe. **(d)** Threshold switching current density in YPtBi as a function of pulse width, where dots are experimental data and solid lines show fitting results given by equation (3).

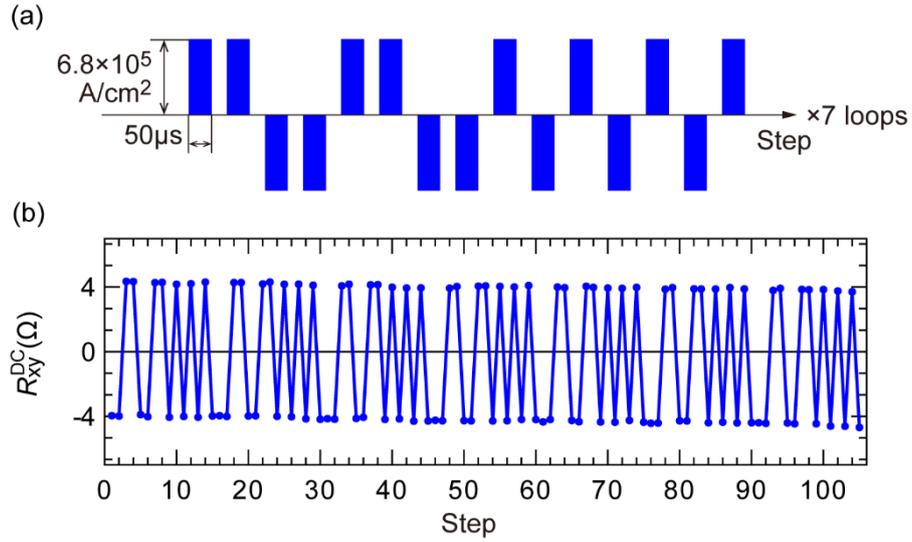

**Figure 9.** Robust SOT magnetization switching in the heterostructure sample whose YPtBi was grown at 300°C with $P_{Ar}$ = 1.0 Pa. (a) Sequence of 105 pulses with the current density in YPtBi of $6.8\times10^5$ Acm$^{-2}$ and the pulse width of 50 μs. (b) Hall resistance data measured under an in-plane bias field of 0.18 kOe.